\documentclass[notitlepage, twocolumn, prl, nobalancelastpage, amssymb, superscriptaddress, showpacs, preprintnumbers, nofootinbib]{revtex4-2}
\usepackage[utf8]{inputenc}
\usepackage[normalem]{ulem}
\usepackage[dvipsnames]{xcolor}
\usepackage{cancel}

\definecolor{red}{rgb}{0.9, 0,0}
\definecolor{cerulean}{rgb}{0., 0.42,0.9}
\definecolor{navy}{rgb}{0.05, 0.05,0.8}

\usepackage[colorlinks]{hyperref}
\hypersetup{
colorlinks = true,
linkcolor = red,
urlcolor  = blue,
citecolor = blue,
anchorcolor = blue
}

\usepackage{amsmath,amsfonts,amsthm, bm, mathtools, latexsym,amssymb} 
\usepackage{hyperref}
\usepackage{subcaption}
\usepackage{physics}
\usepackage[dvipsnames]{xcolor}
\usepackage{enumerate}
\usepackage{slashed}
\usepackage[final]{showlabels}

\usepackage{tikz}
\usepackage{tikz-3dplot}
\usepackage{tikz-feynman,contour}
\usetikzlibrary {positioning}

\def\e{{\epsilon}}

\def\ka{{\kappa}}

\def\a{{\alpha}}

\def\g{{\gamma}}

\def\d{{\delta}}

\def\O{\Omega}

\def\th{\theta}
\def\bz{{\bar z}}

\def\bw{{\bar w}}

\def\CB{{\mathcal B}}

\def\CD{{\mathcal D}}

\def\CH{{\mathcal H}}
\def\CI{{\mathcal I}}

\def\CO{{\mathcal O}}

\def\A{{\mathcal A}}

\def\Tr{{\text{Tr}}}

\newcommand{\tu}{{\tilde{u}}}
\newcommand{\dt}{{\text d}}
\newcommand{\p}{\partial}

\newcommand{\soft}{{\text{\text{soft}}}}

\def\tr{{\tilde{r}}}

\def\CD{{\text{\begin{tiny}CD\end{tiny}}}}
\def\AFG{{\text{\begin{tiny}AFG\end{tiny}}}}

\NewDocumentCommand{\codeword}{v}{%
\texttt{\textcolor{blue}{#1}}%
}

\usepackage{subfiles} 

\begin{document}


\title{From Asymptotically Flat Gravity to Finite Causal Diamonds}
\author{Luca Ciambelli}
\affiliation{Perimeter Institute for Theoretical Physics, 31 Caroline Street North, Waterloo, Ontario, Canada N2L 2Y5}
\author{Temple He}
\affiliation{Walter Burke Institute for Theoretical Physics, California Institute of Technology, Pasadena, CA 91125}
\author{Kathryn M. Zurek}
\affiliation{Walter Burke Institute for Theoretical Physics, California Institute of Technology, Pasadena, CA 91125}

\begin{abstract}
\noindent We demonstrate that the phase space of the soft sector of asymptotically flat gravity in four spacetime dimensions can be identified with that of a spherically symmetric finite casual diamond in Minkowski spacetime. The leading soft graviton mode is geometrically identified with the radial fluctuation of the causal diamond size, while the Goldstone mode involves both the radial fluctuation and its symplectic partner. This allows us to relate the radial fluctuations of the causal diamond with the asymptotic transverse fluctuations parametrized by the soft modes.
\end{abstract}

\maketitle

\preprint{CALT-TH 2025-039}

\maketitle
	
\noindent {\bf Introduction.} 
The analysis of the phase space of gravity in asymptotically flat spacetimes has a rich history (e.g., see \cite{Strominger:2017zoo}). In particular, in the low-energy sector, the phase space is entirely described by boundary modes living on the codimension-2 sphere at $\CI^+_-$, the past horizon of future null infinity $\CI^+$ \cite{Bondi:1962px, Sachs:1962zza, Strominger:2013jfa, He:2014laa}. These boundary modes are precisely the leading soft graviton mode $N$ and its symplectic partner $C$, and this equivalence has led to many fruitful endeavors, including the infrared triangle \cite{He:2014laa, Strominger:2016wns, He:2014cra, He:2015zea}, celestial holography \cite{Kapec:2014opa, Pasterski:2015tva, Pasterski:2016qvg, Kapec:2016jld}, and Carrollian holography \cite{Bagchi:2016bcd, Ciambelli:2018wre, Donnay:2022aba}. These topics are reviewed in \cite{Strominger:2017zoo, Pasterski:2021rjz, Raclariu:2021zjz, Bagchi:2025vri, Nguyen:2025zhg} (see also references therein).

More recently, there has also been a growing interest in the study of finite subregions in a gravitational background, especially concerning causal diamonds and their geometric and thermodynamic properties~\cite{Jacobson:2015hqa, Jacobson:2018ahi, Krishnan:2019ygy, Balasubramanian:2023dpj, Fransen:2025npa, Caminiti:2025hjq}, as well as their quantization~\cite{Verlinde:2019xfb,Banks:2021jwj,Verlinde:2022hhs,Bak:2024kzk,Bub:2024nan,Ciambelli:2025flo,He:2025hag}.  Indeed, for the simple setup involving a spherically symmetric causal diamond in Minkowski spacetime, it was shown in \cite{Bub:2024nan} that the phase space consists of a pair of edge modes living on the codimension-2 bifurcate horizon. One mode is the area $A = 4\pi L^2$ parametrizing the size of the causal diamond, and the other is its symplectic conjugate $\mu$ parametrizing the rate at which the area of the transverse sphere changes along the causal horizon.

\begin{figure}[t]
\centering
\begin{tikzpicture}[scale=0.5]
    \draw[->] (0,-4) -- (0,4) node[above] {$t$};
    \draw[->] (-1,0) -- (4,0) node[right] {$r$};

    \draw[thick, black] (0,3) -- (3,0) -- (0,-3) --  cycle;
    \node at (1,2.8) { $\CH^+$ };
    \node at (1,-2.8) { $\CH^-$ };
    \node at (3,0) [circle, fill, red, inner sep=1.5pt] {};
    \node[below right, red] at (3,0) {$\CB \; (\mu,A)$};

     \draw[thick] (0,3.5) -- (3,3.5);
    \draw[thick] (0,3.4) -- (0,3.6); 
    \draw[dashed] (0,3.4) -- (0,3);
    \draw[thick] (3,3.4) -- (3,3.6);   
    \draw[dashed] (3,3.4) -- (3,0);
    \node at (1.5, 3.8) { $L$ };

    
    \draw[thick, black] (8,5) -- (13,0) -- (8,-5) --  cycle;
    \node at (11,3) { $\CI^+$ };
    \node at (11,-3) { $\CI^-$ };
    \node at (13,0) [circle, fill, inner sep=1.5pt] {};
    \node at (8,5) [circle, fill, inner sep=1.5pt] {};
    \node at (8,-5) [circle, fill, inner sep=1.5pt] {};
    \node[below right] at (13,0) {$i^0$};
    \node[above] at (8,5) {$i^+$};
    \node[below] at (8,-5) {$i^-$};

    \draw[dashed] (13,0) -- (14.5,4);
    \draw[dashed] (13,0) -- (14.5,2);
    \draw[thick, black] (14.5,2) -- (14.5,4);
    \node at (14.5,4) [circle, fill, red, inner sep=1.5pt] {};
    \node[above left, red] at (14.5,4) {$\CI^+_-  \; (C, N)$}; 
 
  \def\A{0.1}       
  \def\lambda{1.0}   
  \def\k{360/\lambda} 
  \draw[blue, domain=0:2, samples=400, smooth] (2,0)
    plot (\x + 9 ,{-1.5 +  \A * sin(\k * \x) });
  \draw[blue, domain=0:2, samples=400, smooth] (2,0)
    plot (\x + 10 ,{-0.3 +  \A * sin(\k * \x) });
  \draw[blue, domain=0:2, samples=400, smooth] (2,0)
    plot (\x + 8.5 ,{1.5 +  \A * sin(\k * \x) });
    
\end{tikzpicture}

\caption{\small We draw the two relevant geometric frameworks. On the left, we have a causal diamond of radius $L$ in a Minkowski background. The edge modes are its area $A$ and its symplectic partner $\mu$, and they parametrize the spherically symmetric perturbations localized at the bifurcate horizon $\CB$. On the right, we have the Penrose diagram of an asymptotically flat spacetime, with the blue ripples indicating fluctuations. We blow up spatial infinity $i^0$ and focus on its future boundary $\CI^+_-$. The fluctuations are parametrized by the leading soft graviton mode $N$ and its symplectic partner $C$, and they are localized at $\CI^+_-$.} \label{fig:sh}
\end{figure}
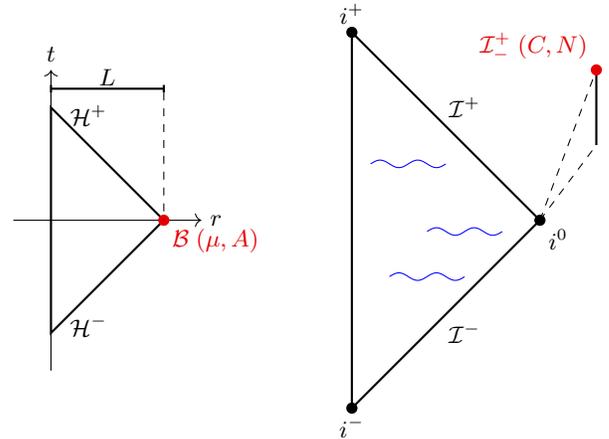

The spherically symmetric causal diamond phase space ostensibly resembles that of asymptotically flat gravity (AFG), and it is therefore tempting to relate the two phase spaces. In this letter, we achieve this to linear order in the leading soft graviton mode $N$ in four dimensions (see Figure~\ref{fig:sh}). By using geometric arguments, we will show that the angle-averaged leading soft graviton mode $\bar N$ is identified with the fluctuating radial mode $\e$ of a causal diamond of radius $L_0$:
\begin{align}\label{intro1}
	\bar N \sim \e .
\end{align}
Physically, this identification captures the fact that area fluctuations are controlled by either $\bar N$, for the celestial sphere at asymptotic infinity, or by $\e$, for a finite causal diamond. We then use the symplectic form to show that the angle-averaged Goldstone mode $\bar C$ involves both $\epsilon$ and $\mu$:
\begin{align}\label{intro2}
	\bar C \sim \mu (L_0+\epsilon).
\end{align} 

The relations Eqs.~\eqref{intro1} and \eqref{intro2} serve to bridge the asymptotic analysis involving soft modes to the experimentally more relevant edge modes of a finite causal diamond.\footnote{The connection between soft and edge modes in gauge theories was recently explored in \cite{Chen:2023tvj, Chen:2024kuq, He:2024ddb, He:2024skc, Araujo-Regado:2024dpr}.} Interestingly, while the soft modes parametrize the transverse fluctuations of the metric due to gravitational radiation, fluctuations in $L \equiv L_0 + \e$ correspond to longitudinal fluctuations of the causal diamond that changes its radius. Part of our goal is to relate these two types of fluctuations to each other, at least for spherically symmetric configurations. By design, the quantity $\e$ is intimately related to potential observables. Therefore, this letter opens up an avenue to further develop the relationship between the asymptotic soft modes, their quantization, and actual experimental observables, ultimately allowing for a deeper understanding of the ideas proposed and explored in \cite{Verlinde:2019xfb, Verlinde:2022hhs, He:2023qha, He:2024vlp}.

\medskip

\noindent{\bf Causal Diamond Phase Space.}
We first review the phase space of a four-dimensional spherically symmetric causal diamond, with radius $L$ and whose past and future horizons are $\CH^\pm$, in Minkowski spacetime. In this section, we will mostly follow the notation and conventions of \cite{Bub:2024nan}, though we will choose to center the causal diamond at the origin, as was done in \cite{Ciambelli:2025flo}. In Gaussian null coordinates, such a causal diamond can be described by the line element\footnote{We slightly deviate from the conventions of \cite{Bub:2024nan} by adding tildes to the $(\tu,\tr)$ coordinates to differentiate them from the coordinates we will later use to describe asymptotically flat spacetimes.}
\begin{align}\label{cd-metric}
\begin{split}
	\dt s^2 &= -2 \ka \tr\, \dt \tu^2 + 2 \,\dt \tu\, \dt \tr + 2\Phi(\tu,\tr)^2 \g_{z\bz} \,\dt z \, \dt \bz \\
	\Phi(\tu,\tr) &\equiv L - \frac{1}{2\ka} e^{\ka \tu + \a} - \tr e^{-\ka \tu - \a},
\end{split}
\end{align}
where $\ka \equiv \ka(L)$ is a spacetime constant known as the inaffinity and only depends on $L$,\footnote{We may view $\ka$ as being the acceleration of a Rindler observer with proper time $\tu$ and whose future Rindler horizon is $\CH^+$.} $\Phi$ the dilaton controlling the size of the transverse sphere, $\a$ an offset in null time $\tu$, and $(z,\bz)$ complex coordinates parametrizing the transverse sphere with round metric $\g_{z\bz} = \frac{2}{(1+z\bz)^2}$. In terms of the retarded null time $u = t- r$, we have
\begin{align}\label{u-tu}
	u = - L + \frac{1}{\ka}e^{\ka \tu + \a}.
\end{align}
The future horizon $\CH^+$ is at $\tr=0$, and the bifurcate horizon $\CB$ is reached by further taking $\tu = \tu_- \to -\infty$. Notice that at the bifurcate horizon, we have $\Phi|_\CB = L$. Furthermore, the top tip of the causal diamond, which we denote as $\CH^+_+$ and is located at $\tu=\tu_+$, is assumed to be a caustic so that $\Phi(\tu_+,0) = 0$. It is straightforward to work out \cite{Bub:2024nan}
\begin{align}\label{uplus}
	\tu_+ = \frac{1}{\ka} \big( \log(2\ka L) - \a \big).
\end{align}

Utilizing the covariant phase space formalism, it was shown in \cite{Bub:2024nan, Fransen:2025npa} that in four spacetime dimensions, the on-shell symplectic form is localized onto the bifurcate horizon $\CB$ and is given by
\begin{align}\label{symp-CD}
\begin{split}
	\O_{\CD} &= \frac{1}{8\pi G} \d \mu \wedge \d A \\
	A &\equiv 4\pi L^2, \qquad \mu \equiv \frac{1}{2} \log \frac{2\p_{\tu}\Phi}{\p_{\tr} \Phi} \bigg|_{\CB} ,
\end{split}
\end{align}
where $G$ is Newton's constant. We can simplify the relationship between $\mu$ and $\Phi$ by observing that Eq.~\eqref{cd-metric} implies
\begin{align}\label{D-Phi}
	\p_{\tu}\Phi\big|_{\CH^+} = - \frac{1}{2} e^{\ka \tu + \a} , \qquad \p_{\tr}\Phi\big|_{\CH^+} = - e^{-\ka\tu-\a},
\end{align}
which in turn means
\begin{align}\label{exp-mu}
	e^\mu = -2 \p_{\tu} \Phi\big|_{\CB} = e^{\ka\tu_- + \a}.
\end{align}
Inverting the symplectic form Eq.~\eqref{symp-CD} yields the bracket
\begin{align}\label{CD-comm}
	\{ \mu , A \} = -8  \pi G  .
\end{align}
We would like to rewrite this bracket in a more useful form. Noting that $A = 4\pi L^2$, we can parametrize the phase space in terms of $L$ and $\mu$. Consider now a classical probability distribution of causal diamonds in phase space given by $\rho(L,\mu)$, such that
\begin{align}
	L_0 \equiv \int \dt L \,\dt \mu\, \rho(L,\mu) L.
\end{align}
Here, $L_0$ is a phase space constant encapsulating the average size of causal diamonds in the distribution.\footnote{Upon canonical quantization, $\rho$ becomes a density matrix $\hat \rho$, and $L_0$ is the expectation value of the length operator $\hat L$ associated to the state $\hat \rho$, so that $L_0 \equiv \Tr(\hat \rho \hat L)$. } We can then isolate the fluctuating mode of the radius and define
\begin{align}\label{eps}
	\e \equiv L - L_0.
\end{align}
Substituting $A = 4\pi (L_0 + \e)^2$ into Eq.~\eqref{CD-comm}, we get
\begin{align}\label{CD-comm2}
\begin{split}
 \{ \mu, \e \} = - \frac{G}{L} .
\end{split}
\end{align}
This is the final form of the bracket involving the degrees of freedom in a causal diamond. Our goal now is to demonstrate that there exists a natural map between the modes in Eq.~\eqref{CD-comm2} to the soft degrees of freedom of asymptotically flat spacetimes. As we will see, this map involves the identification of the area fluctuations in the asymptotic limit.

\medskip

\noindent{\bf Asymptotic Phase Space.}  
We are interested in asymptotically flat gravity in four dimensions, whose metric near $\CI^+$ in Bondi gauge is to leading order in a large-$r$ expansion given by \cite{Bondi:1962px, Sachs:1962zza}
\begin{align}\label{bondi}
\begin{split}
	\dt s^2 &= - \dt u^2 - 2\, \dt u\, \dt r + 2 r^2 \g_{z\bz} \,\dt z\, \dt \bz \\
	&\qquad + \frac{2m_B}{r} \dt u^2 + r C_{zz} \, \dt z^2 + r C_{\bz\bz} \, \dt \bz^2 + \cdots ,
\end{split}
\end{align}
where $u$ again is the retarded null time, $m_B$ the Bondi mass aspect, and $C_{zz}$ the shear. The future and past boundaries of $\CI^+$ are respectively denoted as $\CI^+_\pm$ and correspond to first taking $r \to \infty$ and then $u \to \pm\infty$, respectively. In the absence of matter, the Bondi mass aspect satisfies the constraint equation
\begin{align}\label{mB-eom}
	\p_u m_B = \frac{1}{4} \big( D_z^2 N^{zz} + D_{\bz}^2 N^{\bz\bz} \big) - \frac{1}{4} N_{zz} N^{zz},
\end{align}
where $N_{zz} \equiv \p_u C_{zz}$ is the news tensor capturing the gravitational radiation reaching $\CI^+$, and $D_z$ the $\g$-covariant derivative. 

Using the covariant phase space formalism, the phase space of such asymptotically flat metrics is \cite{Ashtekar:1981bq}
\begin{align}\label{bms-symp}
\begin{split}
	\O_\AFG = - \frac{1}{16\pi G} \int_{\CI^+} \dt u\,\dt^2z\,\g^{z\bz} \d C_{zz} \wedge \d N_{\bz\bz} .
\end{split}
\end{align}
We are interested in isolating the soft (low-energy) sector of the phase space. 
This amounts to studying the phase space associated to the boundary modes
\begin{align}\label{CN-def}
\begin{split}
	-2D_z^2 C(z,\bz)  &\equiv C_{zz}(u,z,\bz)\big|_{\CI^+_-} \\
	D_z^2 N(z,\bz) &\equiv \int_{-\infty}^\infty \dt u \, N_{zz}(u,z,\bz), 
\end{split}
\end{align}
where $N$ is the soft graviton mode associated to the leading gravitational memory effect, and $C$ the Goldstone mode associated with the spontaneously broken supertranslation symmetry. Equivalently, the soft graviton and Goldstone modes in Eq.~\eqref{CN-def} are captured by the shear profile \cite{He:2014laa, Strominger:2016wns, Donnay:2018neh, He:2024vlp}
\begin{align}\label{soft-shear}
\begin{split}
	C_{zz}(u,z,\bz) &= D_z^2 N(z,\bz) \th(u-u_s) - 2 D_z^2 C(z,\bz),
\end{split}
\end{align}
where $\th(u-u_s)$ is the Heaviside step function, and $u_s$ is the location of the shockwave.

Substituting Eq.~\eqref{soft-shear} into the symplectic form Eq.~\eqref{bms-symp}, we get the symplectic form associated to the soft degrees of freedom, namely
\begin{align}\label{bms-soft-symp}
\begin{split}
	\O_\AFG^\soft &= \frac{1}{8\pi G} \int_{S^2} \dt^2z \,\g^{z\bz} D_z^2 \d C \wedge D_\bz^2 \d N ,
\end{split}
\end{align}
where $S^2$ is the celestial sphere. Inverting the symplectic form, we find that the bracket between the soft and Goldstone mode is given by
\begin{align}
	\big\{ D_z^2 C(z,\bz) , D_\bw^2  N(w,\bw) \big\} &= -8\pi  G \g_{z\bz} \d^2(z-w).
\end{align}
Integrating over the transverse directions, we get the bracket
\begin{align}
\begin{split}
	\big\{ C(z,\bz), N(w,\bw) \big\} &= -8 G S\ln|z-w|^2 \\
	S &= \frac{(z-w)(\bz-\bw)}{(1+z\bz)(1+w\bw)}.
\end{split}
\end{align}
It immediately follows by direct computation that
\begin{align}\label{soft-comm}
	\big\{ C(z,\bz) , D_w^2 D_\bw^2 N(w,\bw) \big\} = - 8\pi G\g_{z\bz}\d^2(z-w) .
\end{align}
In writing Eq.~\eqref{soft-comm}, we have modified our phase space to allow for zero modes of $C$ and $D_w^2 D_\bw^2 N$, which in particular implies that we are allowing $N$ to be a singular function on $S^2$.\footnote{We thank Prahar Mitra for discussions related to this point.} 

It will be useful for us to average the above bracket Eq.~\eqref{soft-comm} over the celestial sphere. The result is
\begin{align}\label{soft-avg-comm}
\begin{split}
	\{ \bar C , \bar N \} = - 2 G,
\end{split}
\end{align}
where\footnote{Note that $\bar C,\bar N$ are precisely the zero modes of $C$ and $D_w^2 D_\bw^2 N$, and $\bar N$ does not vanish because $N$ is allowed to be singular. The physical interpretation of $\bar N$ is given in Eq.~\eqref{mB-N}.}
\begin{align}
\begin{split}
	\bar C &\equiv \frac{1}{4\pi} \int_{S^2} \dt^2z\, \g_{z\bz}  C(z,\bz)   \\
	\bar N &\equiv \frac{1}{4\pi} \int_{S^2} \dt^2w \, \g_{w\bw} D_w^2 (D^w)^2 N(w,\bw),\label{Nbar}
\end{split}
\end{align}
and we used the fact
\begin{align}
	\int_{S^2} \dt^2z\,\g_{z\bz} = 4\pi.
\end{align}
We now proceed to demonstrate that there is a natural map that allows us to identify Eq.~\eqref{soft-avg-comm} with Eq.~\eqref{CD-comm2}.

\medskip

\noindent{\bf Connecting the Two Phase Spaces.}
We will determine the relationship between the causal diamond phase space and that of asymptotically flat spacetimes in two steps. First, we will use a geometric argument to relate the angle-averaged leading soft graviton mode $\bar N$ to the length fluctuation mode $\e$. Then, we will use the symplectic form to obtain a relation between the angle-averaged Goldstone mode $\bar C$ and $\mu$. Throughout our analysis, we will only work to linear order in $N$, which is equivalent to only considering spacetime metric fluctuations in the leading low-energy limit, and is also the regime considered in \cite{He:2023qha, He:2024vlp}.\footnote{It would be interesting to understand how the identification between the phase spaces extend beyond our linear approximation.} 

We begin by relating the soft graviton mode $\bar N$ to $\e$. First, recall that the evolution of the induced radial coordinate on the null hypersurface along $u$ given a fixed $(z,\bz)$ is governed asymptotically by the differential equation \cite{Kapec:2016aqd, Ciambelli:2024swv}
\begin{align}\label{a-def}
\begin{split}
	\p_u \big( r(u,z,\bz)^2\big) = 2 m_B(u,z,\bz) - r(u,z,\bz).
\end{split}
\end{align} 
For the case of a Minkowski causal diamond (CD) described above, the Bondi mass vanishes. Thus, the induced radial coordinate, as well as the area of a cut along the null hypersurface, is given by
\begin{align}
\begin{split}
	&\p_u \big(r_{\CD}(u)^2 \big) =  - r_{\CD}(u) \\
	\implies\quad & r_\CD(u) = \frac{1}{2}(L-u) \\
	\implies\quad & A_\CD(u) = 4\pi r_{\CD}(u)^2 = \pi (L-u)^2,
\end{split}
\end{align}
where we determined the integration constant by requiring $r_\CD$ vanishes at the top tip of the causal diamond, i.e., when $u=L$, due to the presence of the caustic. Therefore, for an unperturbed causal diamond with radius $L_0$, the induced radial coordinate at a given $u$ along the future horizon $\CH^+$ is $\frac{1}{2}(L_0 - u)$. It follows that the area variation involving $\e$ (defined in Eq.~\eqref{eps}) is then 
\begin{align}
\begin{split}
	\Delta A_{\CD}(u) &\equiv \pi (L_0 + \e -u)^2  - \pi (L_0 - u)^2 \\
	&= 2\pi ( L_0 - u) \e +\pi \e^2.\label{aCD}
\end{split}
\end{align}
In particular, the rate at which the area variation changes is constant, and is given by
\begin{align}\label{CD-dA}
	\p_u \Delta A_{\CD} = - 2\pi \e.
\end{align}

Next, let us consider the case of asymptotically flat gravity. Solving Eq.~\eqref{a-def} with a nontrivial $m_B$ in the large-$r$ limit, we get the local area element to be
\begin{align}\label{r-soln}
\begin{split}
	a_{\AFG} &\equiv \g_{z\bz} r_{\AFG}(u,z,\bz)^2 \\
	&= \frac{1}{4}\g_{z\bz} (u_0-u)^2  - 2 \g_{z\bz} \int_u^\infty \dt u' \, m_B(u',z,\bz) + \cdots,
\end{split}
\end{align}
where the $\frac{1}{4}\g_{z\bz}(u_0-u)^2$ term is the (divergent) reference area element corresponding to flat spacetime with $m_B = 0$, and $\cdots$ involve further subleading terms in the large-$r$ limit. From Eq.~\eqref{r-soln}, it is clear that the  area variation from the reference local area element only depends on $m_B$, and is given to be
\begin{align}
\begin{split}
	\Delta a_{\AFG}(u,z,\bz) &\equiv a_{\AFG}(u,z,\bz) -\frac{1}{4}\g_{z\bz} (u_0-u)^2 \\
	&= - 2 \g_{z\bz} \int_u^\infty \dt u' \, m_B(u',z,\bz) .
\end{split}
\end{align}
Comparing this equation with Eq.~\eqref{aCD}, we see that the main difference between the finite causal diamond and the asymptotic case is that the area fluctuations are given by the causal diamond size in the former, while they are encoded in the Bondi mass in the latter. This is due to the fact that the size of the causal diamond becomes infinite in the asymptotic limit, and thus it must be renormalized, leading to its fluctuations being controlled by the Bondi mass, the next-to-leading order correction to its size. This is fully expected, as the Bondi mass controls the (de)focusing of null rays parallel to $\mathcal{I}^{+}$ due to energy flowing out of the system.
 
The total area variation is obtained by integrating over the celestial sphere, i.e.,
\begin{align}\label{area-fluc1}
\begin{split}
	\Delta A_{\AFG}(u) &\equiv \int_{S^2} \dt^2z\, \Delta a_{\AFG}(u,z,\bz) \\
	&= - 8 \pi \int_{u}^\infty \dt u'\, \bar m_B(u'),
\end{split}
\end{align}
where we defined the angle-averaged Bondi mass aspect
\begin{align}\label{mB-avg}
\begin{split}
	\bar m_B(u) &\equiv \frac{1}{4\pi} \int_{S^2} \dt^2 z \, \g_{z\bz} m_B(u,z,\bz) .
\end{split}
\end{align}
We can then determine the rate at which the area variation changes to be\footnote{This agrees with $(140)$ of \cite{Ciambelli:2024swv} for a constant cut in $(z,\bz)$. More generally, one gets an expansion scalar as in \cite{Ciambelli:2025mex}.}
\begin{align}
	\p_u \Delta A_{\AFG}(u) &= 8\pi \bar m_B(u).
\end{align}
We are interested in the area variation at the corner, which corresponds to taking $u \to -\infty$. Matching this with the area variation in the causal diamond given in Eq.~\eqref{CD-dA}, we get
\begin{align}\label{eps-mB0}
	 \e = - 4 \bar m_B(-\infty),
\end{align}
where the equality is implicitly understood to mean that the two sides are equal under an isomorphism between the two phase spaces that preserves the brackets.

We now turn to determining $\bar m_B(-\infty)$ in terms of $\bar N,\bar C$. Substituting the shear profile Eq.~\eqref{soft-shear} into Eq.~\eqref{mB-eom}, we obtain
\begin{align}\label{DmB1}
\begin{split}
	\p_u m_B(u,z,\bz) &= \frac{1}{2}(\g^{z\bz})^2 D_z^2 D_\bz^2 N(z,\bz)\d(u-u_s) + \cdots \\
	&= \frac{1}{8} \Box(\Box+2) N(z,\bz)\d(u-u_s) + \cdots,
\end{split}
\end{align}
where $\cdots$ denotes higher order $\CO(N^2)$ terms, and in the last equality we noted the Laplacian is given by
\begin{align}\label{laplacian}
	\Box N \equiv (D^z D_z + D^\bz D_\bz ) N.
\end{align}
Integrating Eq.~\eqref{DmB1} over $\CI^+$ and assuming the boundary condition $m_B|_{\CI^+_+} = 0$, we get to linear order in $N$ 
\begin{align}
\begin{split}
	m_B(u,z,\bz) &= - \frac{1}{8} \Box(\Box+2) N(z,\bz) \big( 1 - \th(u-u_s) \big) .
\end{split}
\end{align}
Performing an angle-averaging as defined in Eq.~\eqref{mB-avg} and evaluating at $\CI^+_-$, we get
\begin{align}\label{mB-N}
\begin{split}
	\bar m_B(-\infty) = - \frac{1}{2} \bar N,
\end{split}
\end{align}
where we used the definition of $\bar N$ given in Eq.~\eqref{Nbar}.
We can now substitute Eq.~\eqref{mB-N} into Eq.~\eqref{eps-mB0} to get, at linear order in $N$, the relationship 
\begin{align}\label{N-eps}
\begin{split}
	\bar N = \frac{1}{2} \e .
\end{split}
\end{align}
This is the first important identification in the matching of phase space variables.

Next, having determined the relation between $\bar N$ and $\e$, we turn to using the symplectic analysis to relate $\bar C$ to the causal diamond modes. Recalling Eq.~\eqref{soft-avg-comm} and substituting in Eq.~\eqref{N-eps}, we get the bracket
\begin{align}\label{C-comm1}
	 \{\bar C,\e \} =  - 4 G.
\end{align}
Further utilizing Eq.~\eqref{CD-comm2}, we get the identification\footnote{We stress that this is an identification at the level of the classical phase space. It would be interesting to study what this identification implies at the quantum level.}
\begin{align}\label{C-Phi}
\begin{split}
	\bar C = 4 \mu L + h(L) ,
\end{split}
\end{align}
where $h$ is an arbitrary phase space function of $L$, and we recall $L=L_0+\epsilon$. The function $h(L)$ arises because the bracket between $\e$ and itself vanishes, and this can be used to our advantage as follows.

From the definition of $\mu$ given in Eq.~\eqref{exp-mu}, we have
\begin{align}\label{mu-explicit}
	\mu = \ka \tu_- + \a,
\end{align}
implying that $\mu$ is formally divergent since $\tu_- \to -\infty$. Therefore, a natural choice for $h$ is to eliminate this coordinate divergence by choosing
\begin{align}\label{h-choice}
	h(L) = - 4 \ka \tu_- L,
\end{align}
so that Eq.~\eqref{C-Phi} becomes
\begin{align}\label{C-alpha}
	\bar C =  4 \a L .
\end{align}
This choice of $h$ given in Eq.~\eqref{h-choice} can be understood by first noting that $\ka \sim L^{-1}$ \cite{Bub:2024nan}. Hence, if we take $L \to \infty$ before taking $\tu \to \tu_-$, we have $\mu \to \a$ since $\ka \to 0$, implying that any remaining phase space degree of freedom present in asymptotically flat gravity is related to $\a$. Our choice of $h$ above then ensures the divergent $\ka \tu_-$ term due to taking the limits in the opposite order drops out. Indeed, this identification fixes the $\bar C = 0$ state to correspond to the $\a=0$ state, and completes our identification between the phase space of a finite Minkowski causal diamond and that of asymptotically flat gravity in four dimensions.\footnote{Of course, in principle we can choose a different identification between $\bar C$ and $\alpha$. However, this does not affect any observables since only changes in $\bar C$ are observable.}

\medskip

\noindent{\bf Discussion.}
In this letter, we provided a precise map in four spacetime dimensions between the angle-averaged soft and Goldstone modes of asymptotically flat gravity and the edge modes of a spherically symmetric causal diamond in Minkowski spacetime. Although the methodology we utilized to obtain this mapping is relatively straightforward, there is a conceptual subtlety that we would like to address.

In some respects, it is rather surprising that such a mapping exists. The causal diamond is a subregion in pure Minkowski spacetime, which is Riemann flat. On the other hand, asymptotically flat spacetimes with nontrivial news and shear are not Riemann flat. Therefore, there does not exist any diffeomorphism relating the two spacetimes, and we should not view the two spacetimes as being equivalent. Nevertheless, to understand why we can identify the two phase spaces, we note that we can view the radial fluctuations of the finite causal diamond size as a result of null shockwaves~\cite{Verlinde:2022hhs,Zhang:2023mkf,Bak:2024kzk}. The relationship of shockwaves to soft modes was recently explored in \cite{He:2023qha, He:2024vlp}, where it was established that the shockwave momentum is
\begin{align}
\begin{split}
	P_-(z,\bz) &= \frac{1}{32\pi G} \Box(\Box+2) N(z,\bz) \\
	&= \frac{1}{8\pi G} D_z^2 (D^z)^2 N(z,\bz).
\end{split}
\end{align}
Using Eq.~\eqref{N-eps}, we see that this in turn implies the angle-averaged momentum $\bar P_-$ is given by the radial fluctuation $\e$ via
\begin{align}\label{shock-N}
\begin{split}
	\bar P_- &\equiv \frac{1}{4\pi} \int_{S^2} \dt^2z \, \g_{z\bz} P_- =  \frac{1}{8\pi G} \bar N = \frac{\e}{16\pi G} .
\end{split}
\end{align}
Therefore, it is more appropriate to view the causal diamond as living in a shockwave background rather than pure Minkowski spacetime~\cite{Verlinde:2019xfb,Verlinde:2022hhs,Zhang:2023mkf,Bak:2024kzk}.  This is a direction we will further explore.

There are a few natural other future directions to explore. First and foremost, one of the major drawbacks in our analysis is that spherical symmetry is imposed on the causal diamond, forcing us to angle-average the soft and Goldstone modes of asymptotically flat gravity. To make contact with more realistic physical systems, we need to relax spherical symmetry and allow for the causal diamond to become deformed. We would then expect to match the resultant causal diamond phase space degrees of freedom with the soft graviton and Goldstone modes $N,C$ without any angle-averaging required. Relaxing spherical symmetry will be crucial for extending our analysis to actual experimental setups sensitive to such length fluctuations, e.g., interferometers, and this is a research avenue that we are currently pursuing.

Finally, we have thus far restricted ourselves to a completely classical analysis. Naturally, we can canonically quantize by replacing the brackets with quantum commutators via the replacement rule $i\{\cdot,\cdot\} \to [\cdot,\cdot]$. However, this is true only to leading order in an $\hbar$ expansion, and there are subtleties involving operator ordering. For instance, since $\a$ and $L$ are conjugate modes and hence do not commute, the identification Eq.~\eqref{C-alpha} is sensible only after an operator ordering is specified. We leave a more complete treatment, which would account for subleading $\hbar$ corrections, for future work.

\medskip

\noindent\textbf{Acknowledgments.}
We would like to thank Mathew Bub, Marc Klinger, Sucheta Majumdar, Prahar Mitra, and Sabrina Pasterski for useful discussions. Research at Perimeter Institute is supported in part by the Government of Canada through the Department of Innovation, Science and Economic Development Canada and by the Province of Ontario through the Ministry of Colleges and Universities. L.C. is supported by the Celestial Holography  Simons collaboration. T.H. and K.Z. are supported by the Heising-Simons Foundation “Observational Signatures of Quantum Gravity” collaboration grant 2021-2817, the U.S. Department of Energy, Office of Science, Office of High Energy Physics, under Award No. DE-SC0011632, and the Walter Burke Institute for Theoretical Physics. K.Z. is also supported by a Simons Investigator award.

\bibliography{references_use.bib}

\end{document}